\documentclass[aps, prx, twocolumn, floatfix, superscriptaddress]{revtex4-2}

\usepackage{amsmath,amssymb,amsfonts,amsbsy}
\usepackage{graphicx}
\usepackage{subfig}
\usepackage{dcolumn}
\usepackage{bm}
\usepackage{multirow}
\usepackage{mathtools}
\usepackage{array}
\usepackage{color}
\usepackage[normalem]{ulem}
\usepackage[per-mode=symbol]{siunitx}
\usepackage{upgreek}
\usepackage{esvect}
\usepackage{booktabs}
\usepackage[utf8]{inputenc}
\usepackage[T1]{fontenc}
\usepackage{lmodern}
\usepackage[pagewise]{lineno}
\usepackage[labelfont=bf]{caption}
\usepackage[figurename=Fig.]{caption}
\usepackage{etoolbox}
\DeclareSIUnit \belm {Bm}
\usepackage{todonotes}
\usepackage{braket}
\usepackage{caption}
\usepackage{float}
\usepackage{tabularx}

\usepackage[hidelinks,bookmarks=false]{hyperref} 
\usepackage[capitalise,nameinlink]{cleveref} 

\newcommand{\qtyerr}[3]{\qty[separate-uncertainty=true,multi-part-units=single]{#1(#2)}{#3}}

\def\@setaltaffiliation{\vspace{-\baselineskip}\def\altaffiliation##1{\@par##1\@addpunct.}\altaffiliationes}
\def\@setaltaffiliation{\vspace{-\baselineskip}\def\altaffiliation##1{\@par##1\@addpunct.}\altaffiliationes}

\let\oldequation\align
\let\oldendequation\endalign

\renewenvironment{align}
  {\linenomathNonumbers\oldequation}
  {\oldendequation\endlinenomath}

\begin{document}

\newcolumntype{L}[1]{>{\raggedright\arraybackslash}p{#1}}
\newcolumntype{C}[1]{>{\centering\arraybackslash}p{#1}}
\newcolumntype{R}[1]{>{\raggedleft\arraybackslash}p{#1}}

\title{Two-qubit logic between distant spins in silicon}

\author{Jurgen~Dijkema}
\thanks{These authors contributed equally to this work}
\author{Xiao~Xue }
\thanks{These authors contributed equally to this work}
    \affiliation{QuTech and Kavli Institute of Nanoscience, Delft University of Technology, Lorentzweg 1, 2628 CJ Delft, Netherlands}

\author{Patrick~Harvey-Collard}
\affiliation{QuTech and Kavli Institute of Nanoscience, Delft University of Technology, Lorentzweg 1, 2628 CJ Delft, Netherlands}

\author{Maximilian~Rimbach-Russ}
\affiliation{QuTech and Kavli Institute of Nanoscience, Delft University of Technology, Lorentzweg 1, 2628 CJ Delft, Netherlands}

\author{Sander~L.~de~Snoo}
\affiliation{QuTech and Kavli Institute of Nanoscience, Delft University of Technology, Lorentzweg 1, 2628 CJ Delft, Netherlands}

\author{Guoji~Zheng}
\affiliation{QuTech and Kavli Institute of Nanoscience, Delft University of Technology, Lorentzweg 1, 2628 CJ Delft, Netherlands}

\author{Amir~Sammak}
\affiliation{QuTech and Netherlands Organization for Applied Scientific Research (TNO), Stieltjesweg 1, 2628 CK Delft, Netherlands}

\author{Giordano~Scappucci}
\author{Lieven~M.~K.~Vandersypen}
\affiliation{QuTech and Kavli Institute of Nanoscience, Delft University of Technology, Lorentzweg 1, 2628 CJ Delft, Netherlands}

\date{\today}

\begin{abstract}
Direct interactions between quantum particles naturally fall off with distance. For future-proof qubit architectures, however, it is important to avail of interaction mechanisms on different length scales. In this work, we utilize a superconducting resonator to facilitate a coherent interaction between two semiconductor spin qubits \qty{250}{\um} apart. This separation is several orders of magnitude larger than for the commonly employed direct interaction mechanisms in this platform. We operate the system in a regime where the resonator mediates a spin-spin coupling through virtual photons. We report anti-phase oscillations of the populations of the two spins with controllable frequency. The observations are consistent with iSWAP oscillations and ten nanosecond entangling operations. These results hold promise for scalable networks of spin qubit modules on a chip. 
\end{abstract}

\maketitle

\section{Introduction}

Solving relevant problems with quantum computers will require millions of error-corrected qubits~\cite{@meter2013}. Efforts across quantum computing platforms based on superconducting qubits, trapped ions and color centers target a modular architecture for overcoming the obstacles to scaling, with modules on separate chips or boards, or even in separate vacuum chambers or refrigerators~\cite{@monroe2016}. For semiconductor spin qubits~\cite{@loss1998,@burkard2023}, the small qubit footprint together with the capabilities of advanced semiconductor manufacturing~\cite{@maurand2016,@zwerver2022} may enable a large-scale modular processor integrated on a single chip~\cite{@vandersypen2017}.  \par

Semiconductor spin qubits are most commonly realized by confining individual electrons or holes in electrostatically defined quantum dots. Nearly all quantum logic demonstrations between such qubits are based on the exchange interaction that arises from wavefunction overlap between charges in neighbouring dots, with typical qubit separations of \qtyrange{100}{200}{\nm}~\cite{@burkard2023,@petta2005}. Monolithic integration of a million-qubit register at a \qty{100}{nm} pitch will face challenges, related to fan-out of control and readout wires. Combining local exchange-based gates and operations between qubits \qtyrange{10}{250}{\um} apart provides a path to on-chip interconnected modules~\cite{@taylor2005,@vandersypen2017}. Various approaches for two-qubit operations over larger distances have been pursued, such as coupling spins via an intermediate quantum dot~\cite{@baart2016,@fedele2021}, capacitive coupling~\cite{@shulman2012}, and shuttling of electrons, propelled either by gate voltages~\cite{@fujita2017,@flentje2017} or by surface acoustic waves~\cite{@jadot2021}. However, the first two methods are still limited to submicron distances and the use of surface acoustic waves faces many practical obstacles, especially in group \MakeUppercase{\romannumeral 4} semiconductors. Electrically controlled shuttling, while being regarded as a promising route, is relatively slow and therefore the coupling distance is constrained by the relevant coherence times. On the other hand, integrating spin qubits with on-chip superconducting resonators using the circuit quantum electrodynamics (cQED) framework provides an elegant way of constructing an on-chip network~\cite{@blais2021,@burkard2020}. With this approach, coupling distances of several hundreds of \unit{\um} are achievable while operations can be just as fast as those based on wavefunction overlap.
\par
In recent years, strong spin-photon coupling \cite{@mi2018,@samkharadze2018,@landig2018} and resonant spin-photon-spin coupling ~\cite{@borjans2019} have both been reported in hybrid dot-resonator devices. 
Among the many ways for constructing distant quantum gates in this architecture (see~\cite{@burkard2020} and references therein), coupling the spins dispersively via virtual photons looks highly promising~\cite{@benito2019,@warren2021}. In this regime, the frequencies of the two qubits are detuned from the resonator frequency and leakage of quantum information into resonator photons is largely suppressed~\cite{@majer2007}. Spin-spin coupling in the dispersive regime has been observed recently in spectroscopy~\cite{@collard2022}, but a two-qubit gate and operation in the time domain remain to be demonstrated. \par

In this work, we demonstrate time-domain control of a dot-resonator-dot system and realize two-qubit logic between distant spin qubits. The two qubits are encoded in single-electron spin states and they are coupled via a 250-\unit{\um}-long superconducting NbTiN on-chip resonator. The resonator is also used for dispersively probing the spin states~\cite{@viennot2015,@mi2018}. First, we demonstrate operations on individual spin qubits at the flopping-mode operating point~\cite{@benito2019b,@croot2020} and characterize the corresponding coherence times. Then, we realize iSWAP oscillations between the two distant spin qubits in the dispersive regime. We study how the oscillation frequency varies with spin-cavity detuning, spin-photon coupling strength, and the frequency detuning between the two spin qubits, and compare the results with theoretical simulations.
\linebreak
\linebreak
\linebreak
\linebreak
\linebreak

\begin{figure*}[htbp] 
\center{\includegraphics[width=\linewidth]{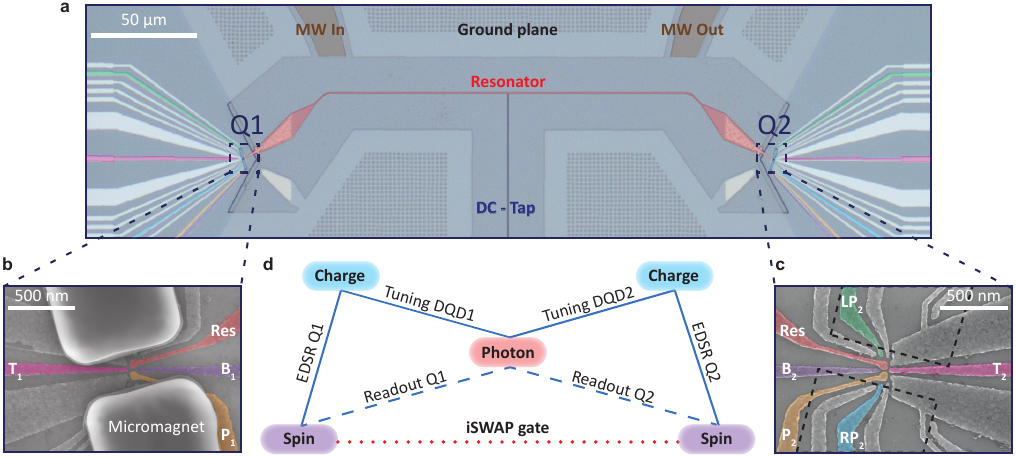}}
\caption{Dot-resonator-dot device. \textbf{a.} False-colored optical image of the device that is used in this work, showing the resonator, ground planes and the gate fan-out of the double quantum dots (DQDs). The microwave (MW) in/out ports are used to probe the transmission through the resonator, and the DC-tap is used to bias the shared top gate Res (in \textbf{b} and \textbf{c}) of the two DQD's. \textbf{b.} Scanning electron microscope (SEM) image of a DQD with the same design as the measured DQD$_1$ showing the gate pattern and the micromagnets on top. \textbf{c.} SEM image of a similar DQD without micromagnets to show the full gate layout of DQD$_2$. The dashed lines outline the micromagnets of the measured device. The false-colored gates in \textbf{b} and \textbf{c} are plunger gates Res and P$_i$, tunnel-barrier gates B$_i$ and T$_i$, and side gates RP$_i$ and LP$_i$. \textbf{d.} Schematic of the distant spin-spin coupling architecture. Blue and red lines indicate dispersive and resonant coupling respectively. Solid lines represent direct couplings, which are the charge-photon electric-dipole coupling and the spin-charge coupling enabled by the micromagnet gradients. These couplings are utilized for probing the DQD charge susceptibility, which allows us to tune the DQD to the degeneracy point, and for electric-dipole spin resonance, respectively. Dashed lines indicate the indirect spin-photon couplings which enable dispersive spin readout. The dotted line indicates a second-order indirect coupling, i.e. the spin-spin coupling mediated by virtual photons, which enables the iSWAP gate.}
\label{fig:device}
\end{figure*}

\section{Device}

The device is fabricated on a $^{28}$Si/SiGe heterostructure and was used in a previous experiment to demonstrate dispersive spin-spin coupling via spectroscopic measurements~\cite{@collard2022}. It contains an on-chip superconducting resonator with an impedance of 3 k$\Omega$~\cite{@samkharadze2016} and a double quantum dot (DQD) at both ends, with gate filters~\cite{@collard2020} (\cref{fig:device}a). The resonator is etched out of a \qtyrange{5}{7}{\nm} thick NbTiN film, and is  \qty{250}{\um} long. Its fundamental halfwave mode, with $\omega_r/2\pi$ = \qty{6.9105}{GHz} and a linewidth of $\kappa_r/2\pi$ = \qty{1.8}{MHz} $\left(Q \approx 3800 \right)$, is used in the experiment for both two-qubit logic and dispersive spin readout. The DQD is defined by a single layer of Al gate electrodes (\cref{fig:device}c). The gates labeled Res are galvanically connected to the resonator. On top of each DQD, a pair of cobalt micromagnets is deposited (\cref{fig:device}b). The device is mounted on a printed-circuit board attached to the mixing chamber of a dilution refrigerator with a base temperature of \qty{8}{mK} (see Appendix B for additional details on the experimental set-up).\par

We accumulate electrons in the DQD$_i$ ($i=1,2$) via plunger gates Res and P$_i$. The interdot tunnel couplings are controlled by the tunnel-barrier gates, labeled T$_i$ and B$_i$. The side gates RP$_i$ and LP$_i$ are used to control the electrochemical potentials and thus the detuning for each DQD. In time-domain experiments, we pulse the detunings via the RP$_i$ gates and drive single qubits by applying microwave bursts through the LP$_i$ gates. The DC voltages on these gates are chosen such that DQD$_1$ and DQD$_2$ are at the degeneracy point between the (1,0)-(0,1) and the (2,3)-(3,2) charge configurations respectively, with both tunnel couplings \qty{\sim 4.8}{GHz} ((m,n) indicates the number of electrons in each DQD). At the charge degeneracy point, the electron is delocalized between the two dots and the resulting charge dipole is maximized, enabling a charge-photon coupling strength of \qty{\sim 192}{MHz} for both DQD's. The magnetic field gradient produced by the micromagnets results in the hybridization of spin and charge states, which allows an indirect electric-dipole interaction between photons and spins (\cref{fig:device}d). The spin-photon coupling can be effectively switched off by pulsing the detuning via the RP$_i$ gate, to an operating point where the electron is tightly localized in a single dot and its electric susceptibility is suppressed. The micromagnets are tilted by $\pm \qty{15}{\degree}$ relative to the double dot axis, which permits tuning the Zeeman energy difference between the two spin qubits by rotating the external magnetic field~\cite{@borjans2019}. In the presence of an external magnetic field $B_{\textrm{ext}}$ of \qty{50}{mT} and at an angle of \qty{4.7}{\degree}, both qubits are set to a frequency of \qty{\sim6.82}{GHz}, slightly detuned from the resonator frequency. The two spins then resonantly interact with each other, mediated by virtual photons in the resonator (\cref{fig:device}d).

\section{Flopping-Mode Qubits}

We first separately examine the individual qubits located at both ends of the resonator. We encode the qubits in two eigenstates at the charge degeneracy point, with $\ket{-,\downarrow}$ as logical $\ket{0}$, and $\alpha \ket{-,\uparrow}+\beta \ket{+,\downarrow}$ as logical $\ket{1}$. Here $\ket{\downarrow}$/$\ket{\uparrow}$ are the spin states, $\ket{-}$/$\ket{+}$ the bonding and antibonding orbitals, and the coefficients $\alpha$/$\beta \gg 1$ are determined by the degree of spin-charge hybridization~\cite{@benito2017,@hu2012} ($\ket{\pm} = \left(\ket{L} \pm \ket{R}\right)/\sqrt{2}$ where $\ket{L}/\ket{R}$ indicate the states where an electron occupies the left/right dot respectively). 
The qubits are manipulated using electric-dipole-spin-resonance (EDSR) enabled by the micromagnets \cite{@pioro2008}. Operation at the charge degeneracy point, chosen to maximize charge-photon and spin-photon coupling strengths, implies that also the EDSR Rabi frequencies are maximized, since they rely on the same matrix element (see also \cref{fig:device}d). This regime is also known as the flopping mode regime~\cite{@croot2020}. Furthermore, since the detuning between the spin and cavity is larger than the spin-photon coupling strength (see below), the spin qubit state is only weakly hybridized with the photons. \par

\begin{figure}[htbp] 
\center{\includegraphics[width=\linewidth]{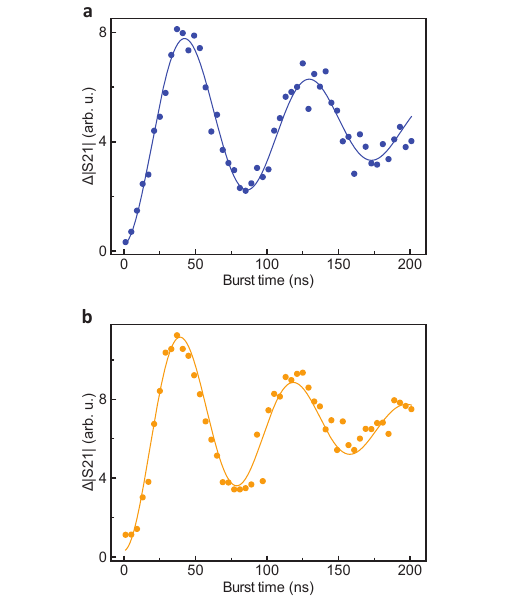}}
\caption{Rabi oscillations for qubit 1 (\textbf{a}) and qubit 2 (\textbf{b}), driven in the flopping-mode regime. The measured transmission through the resonator as a function of applied MW burst time, resonant with qubit 1 and qubit 2 respectively, is shown by the data points. The solid lines represent fits to the data with a damped sinusoid. From the fits, we extract a Rabi decay time $T_2^{\textrm{Rabi}}$ of \qty{100}{\ns} and \qty{120}{ns} for the two qubits. While experimenting on one qubit, the other qubit is parked in the left dot of the corresponding DQD and therefore its coupling to the resonator is effectively switched off. Each data point is averaged for $10^5$ times.}
\label{fig:2}
\end{figure}

\begin{figure*}[htbp] 
\center{\includegraphics[width=\linewidth]{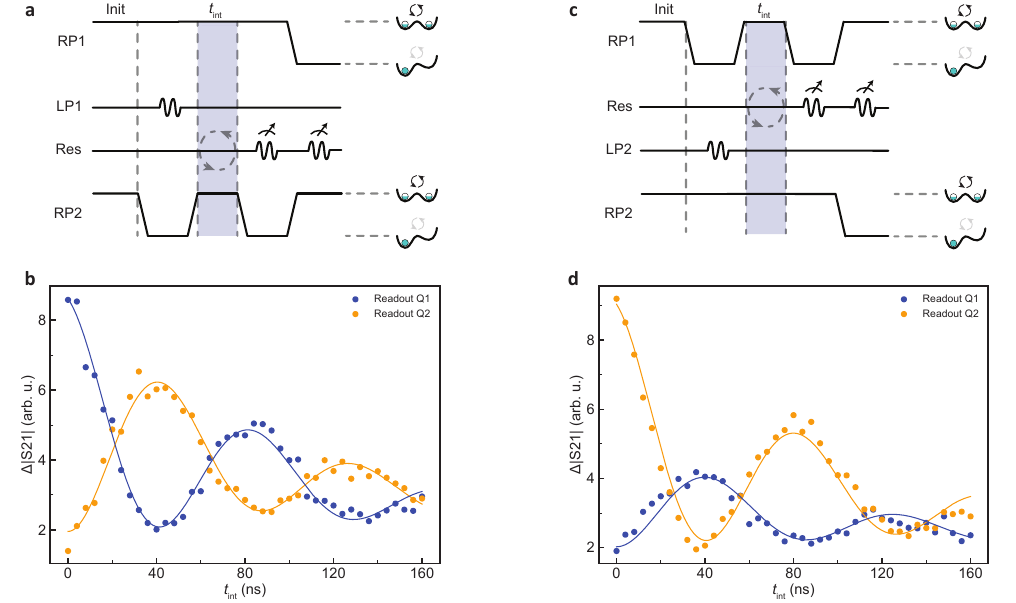}}
\caption{iSWAP oscillations between two distant spin qubits. \textbf{a.} Pulse sequence for the measurement of iSWAP oscillations starting from $\ket{10}$. The two qubits are initialized in $\ket{00}$ by relaxation at the charge degeneracy point. Then qubit 2 is isolated in the left dot for \qty{500}{ns}, during which a $\pi$-pulse is applied to qubit 1. Ten ns later, qubit 2 is pulsed back into interaction, and both spins  interact with each other for $t_{\textrm{int}}$. After the interaction interval, qubit 2 is isolated in its left dot for \qty{1}{\us}, where relaxation is expected to be slower, while qubit 1 is read out using a \qty{400}{ns} long probe tone. Subsequently qubit 1 is isolated in its left dot and qubit 2 is read out using a \qty{400}{ns} probe tone. \textbf{b.} The data points show the measured (change in) transmission, representative of the spin-up population for each spin, starting from $\ket{10}$. Each data point is averaged for $10^6$ times. Here, $g_{s,1(2)}/2\pi \approx $~\qty{21.5}{MHz} and $\Delta_{2s,1(2)}/2\pi  \approx $~\qty{65.5}{MHz}. The solid lines represent fits to the dispersive Hamiltonian model of \cref{eq:H_dispersive}, from which we extract an interaction strength $2J/2\pi$ of \qtyerr{11.6}{0.2}{MHz} \textbf{c.} Pulse sequence similar to \textbf{a} but starting from $\ket{01}$.  \textbf{d.} Similar data and fit as in \textbf{b} but starting from $\ket{01}$. Here we extract $2J/2\pi$ = \qtyerr{11.8}{0.2}{MHz}.}
\label{fig:3}
\end{figure*}

Spin readout is natively achieved in the regime of dispersive spin-photon coupling given that the resonator frequency depends on the electron's spin state. To detect the spin states, we send a microwave probe signal to the resonator at a frequency corresponding to the resonator's frequency with the qubit in $\ket{0}$. In all measurements, the change in microwave transmission (termed transmission for short) relative to this reference value is used as a measure of the $\ket{1}$ population. Due to the limited qubit relaxation times (discussed later), the $\ket{1}$ signal decays within hundreds of nanoseconds. Even though we employ a travelling-wave parametric amplifier (TWPA)~\cite{@macklin2015} to increase the signal-to-noise ratio, signal averaging over many cycles is needed (see Appendix A for a detailed explanation of the readout procedure).\par

For qubit initialization we rely on spontaneous relaxation to the ground state. For this purpose, the short relaxation timescales at the charge degeneracy point are helpful. In practice, a waiting time of \qty{1}{\us} is sufficient for initialization (\qty{\geq 5}{times}~$T_1$). \par

Time-domain single-qubit control of both qubits is illustrated in \cref{fig:2}. We repeatedly initialize the qubit to $\ket{0}$, apply a resonant microwave burst of variable duration through gate LP$_i$, and successively measure both qubits. While we measure one qubit, we pulse the detuning of the other qubit away from the charge degeneracy point so it does not affect the transmission through the resonator. When we plot the average tranmission versus microwave burst time, we observe a damped oscillation, as expected.  \par

Using standard pulse sequences, we next characterize the relaxation and decoherence times of the spin qubits in this system (see the table in~\cref{AP:Decoherence times}). The relaxation ($T_1$) and dephasing ($T_2^*$) times, which are \qtyrange{100}{260}{\ns} and \qtyrange{40}{80}{\ns} respectively, do not match the state-of-the-art spin qubit benchmarks~\cite{@stano2022}. This is expected given the strong spin-charge hybridization at the charge degeneracy point, making the qubits highly sensitive to charge noise~\cite{@benito2019b}. This is the flip side of the faster Rabi oscillations and stronger spin-photon coupling at this working point. Additional contributions to relaxation arise from the Purcell decay induced by the resonator~\cite{@bienfait2016}, and additional decoherence sources include the residual photon population induced by the readout signals~\cite{@yan2018} and intrinsic slow electrical drift in this device.

\begin{figure*}[htbp] 
\center{\includegraphics[width=\linewidth]{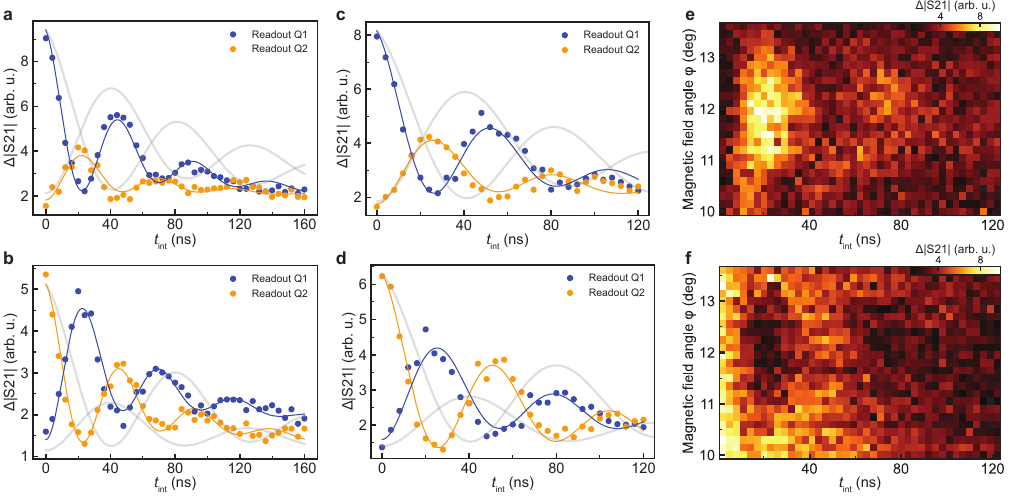}}
\caption{\textbf{a-b.} iSWAP oscillations similar as in \cref{fig:3}b,d. with $g_{s,1(2)}/2\pi $ increased to \qty{\sim 31.9}{MHz}. The fitted oscillation frequencies are now \qtyerr{21.4}{0.3}{MHz} and \qtyerr{21.3}{0.3}{MHz} for \textbf{a} and \textbf{b} respectively. \textbf{c-d}. iSWAP oscillations similar as \cref{fig:3}b,d. with $g_{s,1(2)}/2\pi$ increased to \qty{\sim 31.9}{MHz} and $\Delta_{2s,1(2)}/2\pi$ increased to \qty{\sim 89}{MHz}. Here the fitted frequencies are \qtyerr{18.2}{0.4}{MHz} and \qtyerr{18.7}{0.3}{MHz} for \textbf{c} and \textbf{d} respectively. In \textbf{a-d}, the data points and solid lines represent the measurements and the fit, as in \cref{fig:3}. The faint solid lines reproduce the (rescaled) solid lines from \cref{fig:3}, for comparison. The data points in \textbf{a-d} are averaged for $10^6$ times \textbf{e}. iSWAP oscillations as a function of magnetic field angle $\phi$ with $B_r$ = \qty{49}{mT}, starting from $\ket{01}$ and reading out qubit 1. A chevron pattern is visible, discussed in the main text. \textbf{f}. iSWAP oscillations as function of magnetic field angle similar as \textbf{e} but reading out qubit 2. The data points in \textbf{e-f} are averaged for $10^5$ times.}
\label{fig:4}
\end{figure*}

\section{Two-Qubit Logic}
The spin-spin interaction in the dispersive regime, with both spins resonant with each other but detuned from the resonator, is described by the Tavis-Cummings Hamiltonian~\cite{@tavis1968}. This Hamiltonian describes collective qubit coupling with a resonator and can be simplified to the following dispersive spin-spin coupling Hamiltonian in the rotating-wave approximation~\cite{@burkard2006,@benito2019},

\begin{equation}
    H \approx \hbar J \left( \sigma_{+,1}\sigma_{-,2} + \sigma_{-,1}\sigma_{+,2} \right) \;,
\label{eq:H_dispersive}
\end{equation}

with $J$ the effective spin-spin coupling strength and $\sigma_{\pm,i}$ the usual raising and lowering operators for qubit $i$. The coupling strength is given by

\begin{equation}
    2J = g_{s,1}g_{s,2} \left(\frac{1}{\Delta_{2s,1}} + \frac{1}{\Delta_{2s,2}} \right) \;,
    \label{eq:2J}
\end{equation}
with $g_{s,1}$ and $g_{s,2}$ the spin-photon coupling strength for spin 1 and 2 respectively. $\Delta_{2s,1(2)}$ describes the detuning between the frequency of qubit~1~(2) and the loaded cavity frequency. The interaction with the two charge dipoles dispersively shifts the cavity frequency away from its bare frequency, to \qty{\sim 6.884}{GHz} when both electrons interact with the cavity (both DQDs at charge degeneracy) and to \qty{\sim 6.897}{GHz} when only one electron is coupled (for more detail, see Ref.~\cite{@collard2022}). \par

The Hamiltonian of \cref{eq:H_dispersive} generates iSWAP oscillations between the two spins. We probe this dynamics operating for now in a regime where $g_{s,1(2)}/2\pi \approx$~\qty{21.5}{MHz} and $\Delta_{2s,1(2)}/2\pi \approx $~\qty{65.5}{MHz}. Both spins are initialized to $\ket{0}$ by waiting for \qty{1}{\us} at zero detuning (\cref{fig:3}a,c). Next, one of the spins is prepared in $\ket{1}$, using a calibrated $\pi$-pulse, while the other spin is pulsed away from charge degeneracy in order to effectively decouple it from the cavity. The spins are then allowed to interact with each other by pulsing the second spin back to charge degeneracy, at which point the spins are resonant with each other but still detuned from the (loaded) cavity. After a variable interaction time $t_{\textrm{int}}$, the spins are read out sequentially, once again with the other spin decoupled from the cavity. \par

\cref{fig:3}b (d) shows the measured evolution of both spins starting from $\ket{10}$ ($\ket{01}$). The populations evolve periodically in anti-phase in both experiments, as expected for coherent iSWAP oscillations. The extracted oscillation frequencies are \qty{\sim 11.7}{MHz} for both \cref{fig:3}b,d. The populations of the two spins, separated by more than \qty{200}{\um}, are exchanged in just \qty{\sim 42}{\ns}. A coupling time of \qty{\sim 21}{\ns} is expected to maximally entangle the spins, based on \cref{eq:H_dispersive}. The fidelity of the entangling operation in this regime is numerically estimated to be 83.1\% (see Appendix D). 
The number of visible periods is limited due to the comparatively fast decoherence. Also, because $T_1$ is only a factor \numrange{2}{3} longer than $T_2$, the oscillations are damped asymmetrically towards the ground state. The unequal visibilities of the readout for the two qubits can be attributed to two causes. First, the qubit relaxation times differ, which impacts the signal accumulated during the 400 ns probe interval. Second, the dispersive shift of the resonator frequency has a different magnitude ($\Delta |\textrm{S}21|$) for the two qubits.\par

Next we test whether the measured oscillation frequency varies with control parameters according to \cref{eq:2J}.  First, we increase the spin-photon coupling strength $g_{s,1(2)}/2\pi$ from \qty{\sim 21.5}{MHz} to \qty{\sim 31.9}{MHz} by reducing the tunnel couplings to \qty{\sim 4.35}{GHz}, keeping $B_{ext}$ fixed in order to maintain approximately the same spin-cavity detuning as in \cref{fig:3}. The smaller tunnel coupling decreases the charge-photon detuning, thus increases the charge-induced dispersive shift, and lowers the resonator frequency. Simultaneously, it reduces the spin frequency, due to increased spin-charge admixing. Therefore the spin-cavity detuning stays almost the same as before, $ \Delta_{2s,1(2)}/2\pi \approx$~\qty{63}{MHz}. As expected, we find an increased oscillation frequency of \qty{\sim 21.4}{MHz} (\cref{fig:4}a-b). Next, we increase the spin-cavity detuning to $\Delta_{2s,1(2)}/2\pi \approx $~\qty{89}{MHz} by reducing the external magnetic field $B_{ext}$ from \qtyrange{50}{49}{mT}, while keeping the spin-photon coupling strength the same as in \cref{fig:4}a-b. The angle of the field is re-calibrated and set to \qty{11.5}{\degree} in order to assure the two qubits are still on resonance with each other. In this setting we find the oscillation frequency is reduced from \qty{\sim 21.4}{MHz} to \qty{\sim 18.5}{MHz} (\cref{fig:4}c-d). These fitted oscillation frequencies are slightly different than the predictions from \cref{eq:2J}. We expect the discrepancy will decrease for larger $\Delta_{2s}/g_s$-ratio, i.e. deeply in the dispersive regime (see Appendix D). Consistent with this interpretation, the fitted frequencies are in excellent agreement with the values predicted by a more complete simulation which takes into account the non-zero photon population in the resonator (see Appendix D). \par

One further control knob is given by the frequency detuning between the two qubits, which can be adjusted via the external magnetic field angle $B_\phi$. The oscillation frequency is expected to vary according to $\frac{1}{2\pi} \sqrt{\left(2J\right)^2 + \left(\omega_{Q1} - \omega_{Q2}\right)^2}$, with $\omega_{Qi}$ the angular frequencies of the qubits (see also the spectroscopy data in Ref.~\cite{@collard2022}). As shown in ~\cref{fig:4}e-f, the measurement results display Chevron patterns as a function of the magnetic field angle and the duration of the two-qubit interaction. For $\omega_{Q1}=\omega_{Q2}$ the oscillation is slowest and the $\ket{01}$ and $\ket{10}$ populations are maximally interchanged (maximum contrast). When the magnetic field angle is changed such that $\omega_{Q1}$ differs from $ \omega_{Q2}$, the rotation axis is tilted in the $\ket{01}/\ket{10}$-subspace and we observe accelerated oscillations but with lower contrast, as expected.\par

\begin{figure}[tbp] 
\center{\includegraphics[width=\linewidth]{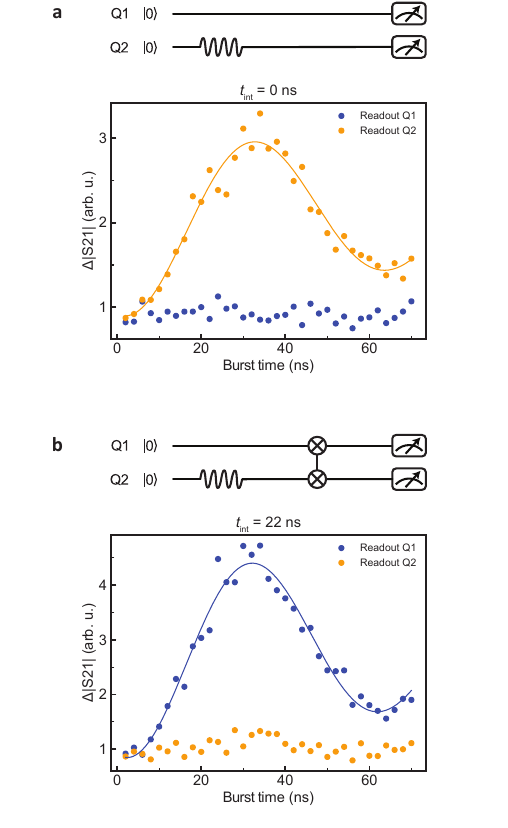}}
\caption{Swapped Rabi-cycle experiment. \textbf{a.} Rabi oscillation of qubit 2 serving as a reference experiment. Starting with both spins in the ground state, a microwave burst of variable duration is applied to qubit 2 followed by readout of this qubit (next the experiment is repeated and qubit 1 is read out). The data points and solid lines represent the measurement and a fit to a sinusoid, as in \cref{fig:2}. Each data point is averaged for $5\cdot10^5$ times.
\textbf{b.} Rabi oscillation of qubit 2 detected through qubit 1. After the microwave burst applied to qubit 2, a calibrated iSWAP gate is executed before readout (performed as in \textbf{b}).}
\label{fig:5}
\end{figure}

Finally, we apply a calibrated iSWAP gate in a practical scenario where the coherent time evolution of one qubit is transferred to the state of the other qubit. First, a reference experiment is executed (\cref{fig:5}a), where we apply a resonant microwave burst of variable duration to qubit 2 in the flopping-mode regime and read out this qubit (orange data points), similar as \cref{fig:2}b. Then we repeat the same measurement but instead read out qubit 1 (blue data points). As seen in the figure, qubit 2 completes a Rabi cycle and qubit 1 remains in the ground state. Next, we perform a similar experiment with an iSWAP gate inserted after the microwave burst, which is expected to map the Rabi oscillation of qubit 2 onto qubit 1 by swapping their populations. As shown in \cref{fig:5}b, the coherence expressed by the Rabi oscillation of qubit 2 is now indeed visible in the final state of qubit 1 (blue data points). Meanwhile qubit 2 arrives in the ground state, which is the initial state of qubit 1 (orange data points).

\section{Conclusion}

Looking ahead, we aim to increase the quality factor of the oscillations and the gate fidelity in several ways. First, the charge qubit linewidth of \qty{\sim 60}{MHz} is much larger than in state-of-the-art devices, where linewidths down to \qty{2.6}{MHz} have been reported~\cite{@mi2017}. Given the admixing of the charge and spin degrees of freedom needed for spin-photon coupling, a narrower charge qubit linewidth simmediately translates to a narrower spin qubit linewidth. Furthermore, the \qty{\sim 30}{MHz} spin-photon coupling strength can be enhanced to at least \qty{\sim 300}{MHz} via both stronger lever arms, a higher resonator impedance and stronger intrinsic or engineered spin-orbit coupling~\cite{@yu2023}. Combining these will allow to work in the deep dispersive regime without compromising on gate speed. With a modest improvement in resonator linewidth, a $>99\%$ two-qubit gate fidelity should be within reach~\cite{@benito2019}. \par

Furthermore, high-fidelity single-qubit operation can be achieved by conventional electric-dipole spin resonance with the electron in a single dot~\cite{@yoneda2017}. Spin readout can be improved by including a third (auxiliary) dot for spin-to-charge conversion based on Pauli spin blockade, enabling rapid and high-fidelity single-shot readout through the resonator~\cite{@zheng2019}. With a dedicated readout resonator or a sensing dot, readout and two-qubit gates can be individually optimized. Finally, this present architecture  allows for exploring two-qubit gates in different regimes, such as the longitudinal coupling regime~\cite{@beaudoin2016, @corrigan2023,@bottcher2022}.\par    

These results mark an important milestone in the effort towards the creation of on-chip networks of spin qubit registers. The increased interaction distance between qubits allows for co-integration of classical electronics and for overcoming the wiring bottleneck. The networked qubit connectivity calls for the design of optimized quantum error correction codes. Moreover, this platform opens up new possibilities in quantum simulation involving both fermionic and bosonic degrees of freedom.

\section*{Acknowledgements}
The authors thank W. Oliver for providing the TWPA, T. Bonsen for helpful insights involving input-output simulations, L. P. Kouwenhoven and his team for access to the NbTiN film deposition, F. Alanis Carrasco for assistance with sample fabrication, L. DiCarlo and his team for access to the $^3$He cryogenic system, O. Benningshof, R. Schouten and R. Vermeulen for technical assistance, and other members of the spin-qubit team at QuTech for useful discussions. This research was supported by the European Union’s Horizon 2020 research and innovation programme under the Grant Agreement No. 951852 (QLSI project), the European Research Council (ERC Synergy Quantum Computer Lab), the Dutch Ministry for Economic Affairs through the allowance for Topconsortia for Knowledge and Innovation (TKI), and the Netherlands Organization for Scientific Research (NWO/OCW) as part of the Frontiers of Nanoscience (NanoFront) program.\par

\textbf{Data and code availability} Data supporting this work and codes used for data processing are available at open data repository 4TU~\cite{Data}.

\textbf{Author contributions} J.D. and X.X. performed the experiment and analyzed the data with help from M.R.-R., M.R-R developed the theory model, P.H.-C. fabricated the device, S.L.d.S. and G.Z. contributed to the preparation of the experiment, G.S. supervised the development of the Si/SiGe heterostructure, grown by A.S. and designed by A.S and G.S., J.D., X.X. and L.M.K.V. conceived the project, L.M.K.V. supervised the project, J.D., X.X., M.R.-R. and L.M.K.V. wrote the manuscript with input from all authors.

\textbf{Competing interests}
The authors declare no competing interests.

\bibliography{library}

\appendix
\onecolumngrid

\clearpage

\section{Calibration and Readout Procedure}

\begin{figure}[htbp] 
\center{\includegraphics[width=\linewidth]{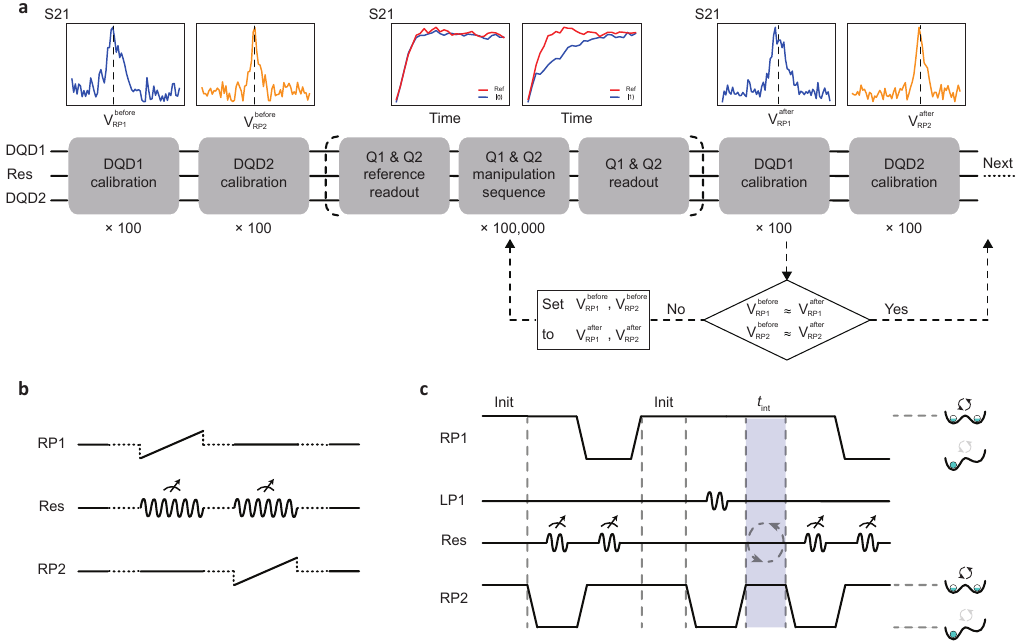}}
\caption{Procedure for the calibration and measurement. \textbf{a.} Workflow of the two-qubit logic measurement. The calibrations for the charge degeneracy points are executed before and after each measurement. The insets on top show example calibration results to find the correct RP$_i$ voltages, where the charge susceptibility is largest. Each calibration scan is averaged for 100 times before moving to the next step. Each single-shot execution of the two-qubit experiment contains reference readout segments followed by a sequence of operations and readout segments for both qubits after the operations. This single-shot experiment is executed 100,000 times before the next round of calibrations. If the calibrations before and after suggest that no drift or only mild drift took place, the result of the 100,000 single-shot is accepted. Otherwise the data is retaken. Next, a control parameter is changed, for instance $t_{\textrm{int}}$, and another round of 100,000 repetitions is executed. After experiments have been completed for all values of the control parameter, the entire procedure is repeated. The final result is analyzed based on 10 such repetitions, thus on 1,000,000 averages. The averaged readout traces are divided into segments of 32 ns, each of which is processed with a fast Fourier transform. The signal persists for a few 100 ns, so we add up the FFT of 13 consecutive segments. The result is shown in the top-middle inset. The enclosed area between the reference readout trace and the readout trace after the sequence of operations on the qubits is recorded as $\Delta|S21|$ in the plots in the main text. \textbf{b.} Pulses used for the calibration of the RP$_i$ voltages of the two qubits. The RP$_i$ gate voltage is swept linearly for 4 $\mu$s in a range of $\pm$11 mV, while the resonator is being probed with a continuous wave of equal duration. The probe frequency is chosen to correspond to the resonance when the DQD is coupled to the resonator so the transmission signal is reduced when the electron is parked in a single dot and is at is maximum value when it is at the charge degeneracy point. \textbf{c.} Complete pulse sequence for measurements of two-qubit iSWAP oscillations, which is plotted in a simplified version in \cref{fig:3}. Before the operations for qubit manipulation, each qubit is measured in its ground state to record the reference readout traces. The integration time is equal to that of the readout traces after the qubit manipulation.}
\label{fig:10}
\end{figure}

\subsection*{Calibration}
As described in Ref.~\cite{@collard2022}, the device undergoes slow electric drift in experiment, which is reflected by the instability of the RP$_i$ and LP$_i$ voltages at the inter-dot transition point. To mitigate the effect of the drift, we implement automated calibration of the RP$_1$ and RP$_2$ voltages for the (1,0)-(0,1) and (3,2)-(2,3) transitions for DQD$_1$ and DQD$_2$ respectively. Note that the voltages applied to LP$_1$ and LP$_2$ are not calibrated, as any drift can be compensated by resetting the RP$_1$ and RP$_2$ voltages.  \par
The sequence of operations for the two-qubit experiments is illustrated in \cref{fig:10}.a. We first apply continuous wave to probe the resonator transmission while linearly sweeping the RP$_1$ gate to calibrate the voltage required for the charge degeneracy point for DQD$_1$, followed by a similar calibration for DQD$_2$~\cref{fig:10}.b. After these calibrations, the RP$_1$ and RP$_2$ voltages are set to the new values and recorded as $V_{\textrm{RP}_1}^{\textrm{before}}$ and $V_{\textrm{RP}_2}^{\textrm{before}}$. Then we apply pulses to control the two-qubit experiment, which includes a reference measurement to extract the readout signal for the $\ket{0}$ state of both qubits and the actual measurement after qubit manipulation~(\cref{fig:10}.c). The integrated difference between the two readout traces is the measurement result and is shown as $\Delta|S21|$ in the plots in the main text. Here, the reference traces help to additionally compensate for drift. After the two-qubit experiment, the RP$_1$ and RP$_2$ voltages for the charge degeneracy points, $V_{\textrm{RP}_1}^{\textrm{after}}$ and $V_{\textrm{RP}_2}^{\textrm{after}}$, are calibrated and compared to $V_{\textrm{RP}_1}^{\textrm{before}}$ and $V_{\textrm{RP}_2}^{\textrm{before}}$. If their differences are both smaller than 0.3 mV, we proceed to the sequence for the next setting. Otherwise, the result is abandoned and the same measurement is retaken. It is important to note that the RP$_i$ gates are located to the side of the dots and thus they have much smaller lever-arms compared to the top plunger gates. In a similar device, the lever-arm of a RP$_i$ gate is measured to be 30 $\mu$eV/mV for the right dot and its lever-arm for the left dot is measured to be 17 $\mu$eV/mV. Thus, the lever-arm for the inter-dot detuning is only 13 $\mu$eV/mV.\par

For single-qubit experiments, the workflow is similar except that the calibration and reference readout are executed only for the 
qubit under investigation while the other qubit is parked in the left dot. \par

\subsection*{Readout}

In the regime of dispersive spin-photon coupling, the frequency of the resonator depends on the state of the qubit. Therefore, the qubit state can be detected by probing the transmission of a readout tone through the resonator. The readout tone is generated at the frequency of the resonator when the qubit is in the $\ket{0}$ state. Thus, a high transmission signal amplitude corresponds to the $\ket{0}$ state whereas a low transmission signal amplitude corresponds to the $\ket{1}$ state. The two states can also be distinguished by the phase difference in the signals but we choose to use the amplitude for readout for its higher sensitivity in our operating regime.\par
Because of the magnetic field and the coupling to the spin, the linewidth of the resonator increases to \qty{\sim 5}{MHz} , corresponding to a response time of \qty{\sim200}{ns}, comparable to the $T_1$'s of the qubits. Given the limited signal-to-noise ratio obtained when averaging for a few 100 ns, single-shot qubit readout does not yield a meaningful fidelity. To enhance the signal-to-noise ratio, we average over many readout traces that have been down-modulated to \qty{50}{MHz}, and then apply a fast Fourier transform to the averaged signal to extract its amplitude and phase.\par

\clearpage
\section{Measurement Setup}
\begin{figure*}[htbp] 
\center{\includegraphics[width=\linewidth]{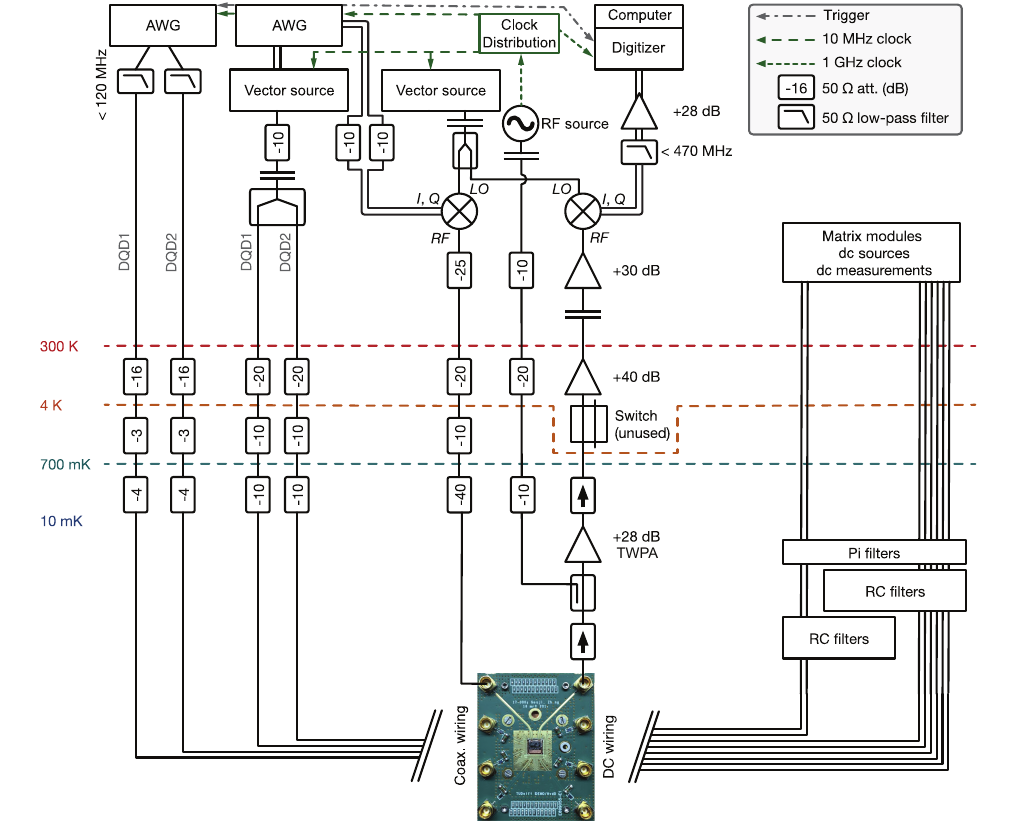}}
\caption{Schematic overview of the measurement setup used in the experiment.  } 
\label{fig:meas_setup}
\end{figure*}

\subsection*{Resonator circuitry} 
An AWG (Tektronix AWG5014C) generates the I,Q signals for the resonator probe tones using a sample rate of \qty{1}{GS/s}. The I,Q signals are attenuated by \qty{10}{dB} to enhance the voltage resolution allowing for improved calibration of the I and Q amplitudes, compensating for possible asymmetries in the up-converting IQ mixer (Marki M1-0307LXP). The output of a vector source (Keysight E8267D) is splitted and provides the LO tone for this mixer as well as for the down-converting mixer. The RF output of the up-converting mixer then reaches the cryostat (Oxford Instruments Triton 400) and connects to the `MW in' resonator feedline on the chip (\cref{fig:device}) after various attenuation stages. \par
\begin{figure*}[htpb] 
\center{\includegraphics[width=\linewidth]{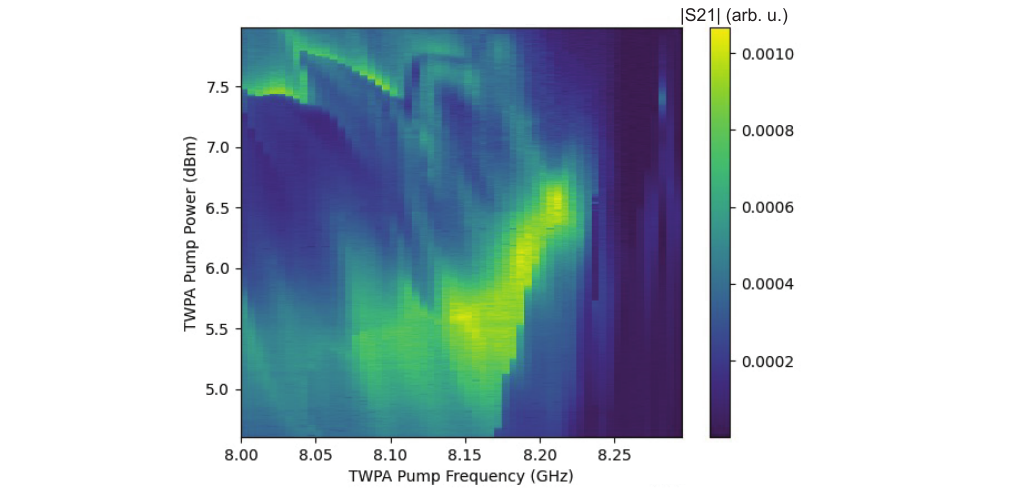}}
\caption{TWPA calibration scan. In order to find the optimal operation point of the TWPA, a calibration scan is executed. The probe frequency is set to the resonator bare frequency of $\omega_r/2\pi$ = 6.9015 GHz, with both electrons parked in single dots, thus decoupled from the resonator. Next, we sweep the RF frequency and RF power of the TWPA pump tone to map out the gain profile. Based on this measurement we choose the set point of the TWPA, which is a power of 5.17 dBm and a frequency of 8.086 GHz. These settings are used for every experiment reported in this work.}
\label{fig:TWPA}
\end{figure*}

The returning signal from the `MW out' feedline is then amplified by a TWPA, which has been configured for maximum gain (\cref{fig:TWPA}), after passing through an isolator (QuinStar 0xE89) and a directional coupler through which the TWPA pump tone enters. After passing another isolator, the signal is amplified at the 4K stage (Low Noise Factory LNF-LNC4-8A) and at room temperature (Miteq AFS3 10-ULN-R) after which it is down-converted to a signal of \qty{50}{MHz}. The signal is then filtered by a \qty{< 470}{MHz} low-pass filter before it is amplified once more (Stanford Research Systems SRS445A) and digitized (AlazarTech ATS9870) with a sample rate of \qty{1}{GS/s}. \par 
The RF source that generates the TWPA pump tone also supplies the reference clock signal to an in-house developed RF reference distribution unit. This unit shares the RF reference signal with the AWG's, vector sources and the digitizer in the setup to synchronize all clocks.

\subsection*{DQD-gates circuitry}
The MW-bursts used to drive the qubits in the experiment are generated by using the internal IQ modulation of the vector source, where an AWG supplies the I,Q signals. The output of the vector source is splitted and connected to the LP$_i$ gates (\cref{fig:device}) after attenuation at the various stages. A separate AWG is used to generate the voltage pulses that allow us to quickly tune the DQD's into and out of the charge degeneracy point. The two lines from the AWG connect to the RP$_i$ gates of both DQD's (\cref{fig:device}) after attenuation at several stages. Home-built voltage sources used to DC-bias the DQD-gates are mounted in IVVI racks and are connected to various home-built matrix modules. The DC lines breaking out from the modules are filtered using home-built Pi filters and subsequently $RC$-filters before reaching the chip in the cryostat.

\clearpage
\section{Parameter table}
\label{AP:Decoherence times}

\begin{table}[h]
\centering
\caption{Parameters used for the various measurements. ``Measured'' indicates a parameter is extracted directly from experiment. ``Estimated'' means the parameter is estimated based on modeling and on other parameters that are directly measured.}
\begin{tabular}{L{3.5cm} C{3cm} R{2cm} R{2cm} R{4cm} }
    \toprule
         \textbf{General} &\textbf{Determination}& {\textbf{Qubit 1} } & {\textbf{Qubit 2} } & \textbf{Other} \\

    \midrule
    $\omega_r/2\pi$ &measured& && 6.9105 GHz \\
    $g_c/2\pi$ & estimated &192 MHz&192 MHz&  \\
    $\Delta B_x$ & estimated &42 mT &42 mT \\
    $T_1$  &measured& 200-260 ns   & 100-130 ns  \\
    $T_2^*$  &measured& 60-80 ns    & 40-60 ns  \\
    $T_2^H$  &measured& 140-160 ns   & 70-90 ns \\
    $T_2^\textrm{Rabi}$  &measured& 100-110 ns   & 100-110 ns \\
    \midrule
    \textbf{Fig. 2} &   & \\
    \midrule
        
    $\chi_c/2\pi$  & measured& 13.5 MHz  & 13.5 MHz  \\
    $g_s/2\pi$  & estimated& 21.5 MHz   & 21.5 MHz  \\
    $\Delta_{2s}/2\pi$  &measured&  65.5 MHz   & 65.5 MHz  \\
    $t_c/h$  &estimated& 4.8 GHz   & 4.8 GHz \\
    $\omega_q/2\pi$  &measured& 6.818 GHz   & 6.818 GHz \\
    \midrule
    \textbf{Fig. 3b,d} &   & \\
    \midrule
    $\chi_c/2\pi$  & measured& 13.5 MHz  & 13.5 MHz  \\
    $g_s/2\pi$  & estimated& 21.5 MHz   & 21.5 MHz  \\
    $\Delta_{2s}/2\pi$  &measured& 65.5 MHz   & 65.5 MHz  \\
    $t_c/h$  &estimated& 4.8 GHz   & 4.8 GHz \\
    $\omega_q/2\pi$  &measured& 6.818 GHz   & 6.818 GHz \\
    $2J/2\pi$              &  measured & & & init $\ket{10}$: \qtyerr{11.6}{0.2}{MHz} \\
             & & & & init $\ket{01}$: \qtyerr{11.8}{0.2}{MHz} \\
            
    \midrule
    \textbf{Fig. 4a-b   \newline Fig. 5} &   & \\ 
    \midrule
    $\chi_c/2\pi$  & measured& 20.5 MHz  & 20.5 MHz  \\
    $g_s/2\pi$  & estimated& 31.9 MHz   & 31.9 MHz  \\
    $\Delta_{2s}/2\pi$  &measured& 63 MHz    & 63 MHz  \\
    $t_c/h$  &estimated& 4.35 GHz   & 4.35 GHz \\
    $\omega_q/2\pi$  &measured& 6.807 GHz   & 6.807 GHz \\
    $2J/2\pi$             &measured   & & & init $\ket{10}$: \qtyerr{21.4}{0.3}{MHz} \\
            & & & & init $\ket{01}$: \qtyerr{21.3}{0.3}{MHz} \\
    \midrule
    \textbf{Fig. 4c-d} \newline \textbf{Fig. 4e-f} at $\phi$ = \qty{11.7}{\degree} &   & \\ 
    \midrule
    $\chi_c/2\pi$  & measured& 20.5 MHz  & 20.5 MHz  \\
    $g_s/2\pi$  & estimated& 31.9 MHz   & 31.9 MHz  \\
    $\Delta_{2s}/2\pi$  &measured& 89 MHz    & 89 MHz  \\
    $t_c/h$  &estimated& 4.35 GHz   & 4.35 GHz \\
    $\omega_q/2\pi$  &measured& 6.78 GHz   & 6.78 GHz \\
    $2J/2\pi$              & measured  & & & init $\ket{10}$: \qtyerr{18.2}{0.4}{MHz} \\
            & & & & init $\ket{01}$: \qtyerr{18.7}{0.3}{MHz} \\
    \bottomrule

\label{tab:parameter overview}    
\end{tabular}    
\end{table}

\clearpage
\section{Simulations}
The system presented in this study consists of two flopping-mode qubits, qubits encoded in individual electron spins confined inside a double quantum dot. Each electron spin is coupled to a single mode of a joint superconductor resonator. This composite system is well described by the following Hamiltonian
\begin{align}
    H=H_\text{res} + H_\text{DQD,1} + H_\text{DQD,2} + H_\text{int,1} + H_\text{int,2}.
    \label{eq:HamTot}
\end{align}
The first term $H_\text{res}=\hbar \omega_r a^\dagger a$ describes photons inside the resonator, where $a^\dagger$ ($a$) creates (annihilates) a photon with angular frequency $\omega_r$. 
The second (third) term describes the dynamics of qubit 1 (2). Each flopping-mode qubit is modelled as a 4-level system~\cite{@benito2019,@collard2022}
\begin{align}
    H_\text{DQD,1(2)} = \frac{1}{2}\left[ \epsilon_{1(2)}\tau_{z,1(2)} + 2t_{c,1(2)} \tau_{x,1(2)} + ( \boldsymbol{h}_{1(2)} + \boldsymbol{\Delta h}_{1(2)} \tau_{z,1(2)}/2)\cdot \boldsymbol{\sigma}_{1(2)}  \right]
\end{align}
with $\boldsymbol{\sigma}_{1(2)}=(\sigma_{x,1(2)},\sigma_{y,1(2)},\sigma_{z,1(2)})^T$. Here, $\tau_{\xi,1(2)}$ and $\sigma_{\xi,1(2)}$ with $\xi=x,y,z$ are Pauli matrices describing the position and spin degree of freedom of the electron in DQD 1 (2). The parameter $\epsilon_{1(2)}$ denotes the energy detuning and $t_{c,1(2)}$ the tunnel coupling between the left and right quantum dot. The effect of the global and micromagnet-induced magnetic field in DQD 1 (2) is described by $\boldsymbol{h}_{1(2)}=\mu_B g_e (B_{L,1(2)}+B_{R,1(2)})/2$ and $\boldsymbol{\Delta h}_{1(2)}=\mu_B g_e (B_{L,1(2)}-B_{R,1(2)})$, with $g_e=2$ being the Lande g-factor of an electron in silicon and $\mu_B$ Bohr's magneton, giving rise to a hybridization between spin and position (charge). A detailed characterisation of the parameters is given in Ref.~\cite{@collard2022}. 

The last terms in Hamiltonian~\eqref{eq:HamTot} describe the charge-photon interaction
\begin{align}
    H_\text{int,1(2)} = g_{c,1(2)} \tau_{z,1(2)}(a^\dagger + a)
\end{align}
with coupling strength $g_{c,1(2)}$. The spin-photon interaction is mediated via the charge degree of freedom.

\subsection*{Reduced models}
Since running the full system is computationally expensive, we eliminate the charge degree of freedom in the limit $\text{max}(g_{c1(2)},|\boldsymbol{\Delta h}_{1(2)}|)\ll 2t_{c,1(2)}$ using standard block diagonalization methods~\cite{@benito2017,@benito2019b,Pymablock}. The reduced Hamiltonians at $\epsilon_{1(2)}=0$ then read
\begin{align}
    H_\text{res} &\approx\hbar (\omega_r-\chi_{1}-\chi_{2}) a^\dagger a,\\
    H_\text{DQD,1(2)} &\approx \frac{1}{2} \omega_{Q1(2)}\sigma_{z,1(2)},\\
    H_\text{int,1(2)} &\approx g_{s,1(2)} (a^\dagger + a)\sigma_{x,1(2)},
    \label{eq:qDickeModel}
\end{align}
where $\hbar\chi_{1(2)} = g_{c,1(2)}^2\left(\frac{1}{2t_{c,1(2)}-\hbar\omega_r}+\frac{1}{2t_{c,1(2)}+\hbar\omega_r}\right)$ is the charge dispersive shift~\cite{kohlerDispersiveReadoutUniversal2018} and $g_{s,1(2)}$ is the spin-photon coupling. The resulting full system Hamiltonian is identical to the 2-qubit Dicke model, a quantum Rabi model with two qubits coupled to a common resonator mode. The Tavis-Cummings Hamiltonian in the main text follows from Hamiltonian Eq.~\eqref{eq:qDickeModel} under the rotating frame approximation $(a^\dagger + a)\sigma_{x,1(2)}\rightarrow(a^\dagger \sigma_{-,1(2)} + a\sigma_{+,1(2)})$ with $\sigma_{\pm,1(2)}=\sigma_{x,1(2)}\pm i \sigma_{y,1(2)}$.

In the limit $g_{s,1(2)}\ll \hbar|\Delta_{Q1(2)}|$ with $\Delta_{Q1(2)} = \omega_r-\chi_{1}-\chi_{2}-\omega_{Q1(2)}$, the so-called dispersive regime, we can further eliminate the photonic degree of freedom. The final Hamiltonian then reads 
\begin{align}
    H_\text{disp} = \frac{\hbar}{2} \omega_{Q1}\sigma_{z,1} +\frac{\hbar}{2} \omega_{Q2}\sigma_{z,2} + 2J (\sigma_{x,1}\sigma_{x,2}+\sigma_{y,1}\sigma_{y,2}),
    \label{eq:qXYModel}
\end{align}
with the interaction strength $J=g_{s,1}g_{s,2} \left(\frac{1}{\hbar\Delta_{Q1}}+\frac{1}{\hbar\Delta_{Q2}}\right)$. Note that Eq.~\eqref{eq:qXYModel} is identical to Eq.~(1) of the main text.

\subsection*{Numerical simulations and noise models}
The dynamics of the system can be computed by solving the Schrödinger equation
\begin{align}
    i\hbar\frac{d}{dt}\psi=H(t) \psi,
\end{align}
where $\psi$ is the quantum state. Additionally, each qubit is subject to relaxation and dephasing channels, the resonator is subject to photon decay, and the system parameters are subject to low-frequency fluctuations.

Qubit relaxation $\gamma_{r,Q1(2)}$, qubit dephasing $\gamma_{\phi,Q1(2)}$, and photon decay $\kappa$ are introduced as Markovian processes within the Lindblad formalism by solving the resulting master equation
\begin{align}
    \frac{d}{dt}\rho = -\frac{i}{\hbar} (H\rho -\rho H) + \sum_i \mathcal{D}_i(\rho),
\end{align}
where $\rho$ is the density matrix, $\mathcal{D}_i(\rho)=\gamma_i (2L_i\rho L_i^\dagger -L_i^\dagger L_i \rho  - \rho L_i^\dagger L_i)/2$ is the dissipation operator. Explicitly, we use the following channels
\begin{alignat}{3}
    \gamma &= \gamma_{r,Q1(2)}\quad && L_{r,Q1(2)} &&= \sigma_{-,1(2)} \\
    \gamma &= \gamma_{\phi,Q1(2)}\quad && L_{\phi,Q1(2)} &&= \sigma_{z,1(2)} \\
    \gamma &= \kappa && \quad\quad L_\kappa &&= a \;.
\end{alignat}

\subsection*{Readout model}
We model the readout via the resonator within the input-output framework in the linear response regime~\cite{combesSLHFrameworkModeling2017}. The measured output signal is then given in the Hilbert-Schmidt or Liouville space by~\cite{bonsenProbingJaynesCummingsLadder2023}
\begin{align}
    S_{21,Q1(2)} = i\hbar\sqrt{\kappa_\text{in}\kappa_\text{out}}\bra{I_n}(a\otimes I_n)(-\mathcal{H}_{Q1(2)} - \mathcal{L} - i\hbar\omega_\text{read}I_{n^2})^{-1}(a^\dagger\otimes I_n - I_n \otimes a^\star)\ket{\rho_\text{final}}.
\end{align}
Here, $I_n$ is an identity matrix of dimension $n=\text{dim}(H)$, $\mathcal{H}_{Q1(2)}=-i(H\otimes I_n - I_n \otimes H^\star)$ is the full Hamiltonian with $g_{s,2(1)}=0$ and $\chi_{2(1)}=0$, and $\mathcal{L}=\sum_i \gamma_i (2L_i\otimes L_i-I_n\otimes (L_i^T L_i)^T-(L_i^T L_i)\otimes I_n)/2$ is the Lindbladian of the system. The vector $\ket{A}$ is the vectorized form of the matrix $A$, $\rho_\text{final}$ is the final density matrix, $\omega_\text{read}$ is the readout frequency, and $\kappa_\text{in(out)}$ is the coupling rate between the resonator and the input (output) line.

\subsection*{Fitting procedure}
\begin{figure*}[htbp] 
\center{\includegraphics[width=\linewidth]{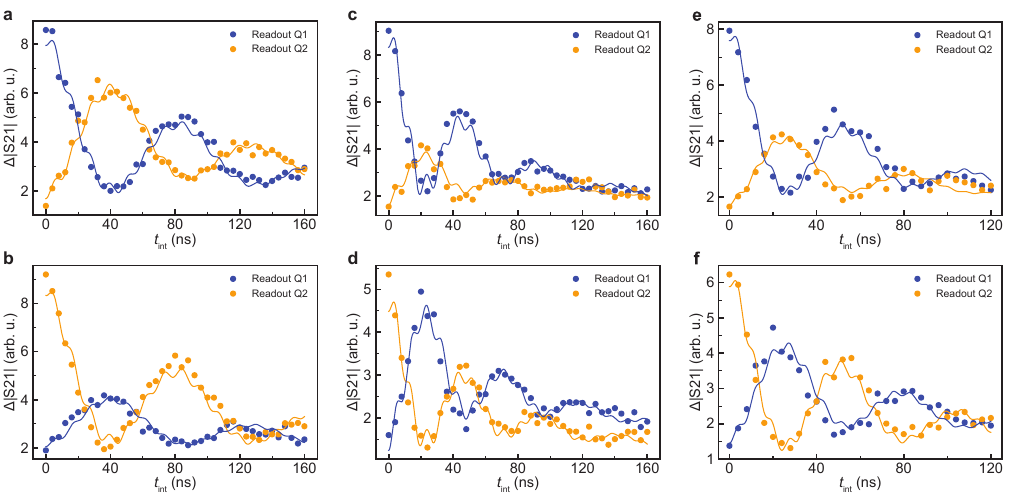}}
\caption{Two-qubit logic data fitted with the full model. The rapid oscillations arising from the fits are due to coherent energy exchange with the resonator via residual vacuum Rabi oscillations. The fitted spin-photon coupling strength $g_s/2\pi$, the spin-spin interaction strength $2J/2\pi$ and the estimated fidelities for the $\sqrt{\text{iSWAP}}$ gate $F_{\sqrt{\text{iSWAP}}}$ and the iSWAP gate $F_{\text{iSWAP}}$ are given hereby. \textbf{a.} $g_s/2\pi$ = 20.9 MHz, $2J/2\pi$ = 11.5 MHz, $F_{\sqrt{\text{iSWAP}}}$ = 82.2\%, $F_{\text{iSWAP}}$ = 67.4\%. \textbf{b.}  $g_s/2\pi$ = 21.0 MHz, $2J/2\pi$ = 11.6 MHz, $F_{\sqrt{\text{iSWAP}}}$ = 84.0\%, $F_{\text{iSWAP}}$ = 72.2\%. \textbf{c.} $g_s/2\pi$ = 29.7 MHz, $2J/2\pi$ = 21.2 MHz, $F_{\sqrt{\text{iSWAP}}}$ = 76.5\%, $F_{\text{iSWAP}}$ = 62.0\%. \textbf{d.} $g_s/2\pi$ = 29.4 MHz, $2J/2\pi$ = 20.9 MHz, $F_{\sqrt{\text{iSWAP}}}$ = 77.5\%, $F_{\text{iSWAP}}$ = 63.2\%. \textbf{e.} $g_s/2\pi$ = 31.0 MHz, $2J/2\pi$ = 18.0 MHz, $F_{\sqrt{\text{iSWAP}}}$ = 80.6\%, $F_{\text{iSWAP}}$ = 66.2\%. \textbf{f.} $g_s/2\pi$ = 31.6 MHz, $2J/2\pi$ = 18.7 MHz, $F_{\sqrt{\text{iSWAP}}}$ = 82.5\%, $F_{\text{iSWAP}}$ = 70.3\%. The estimated  $\sqrt{\text{iSWAP}}$ gate fidelity reported in the main text is an average over the numbers from \textbf{a} and \textbf{b}.}
\label{fig:fullmodel_fits}
\end{figure*}

We use the following fitting procedures. We fit the qubit relaxation and dephasing times, $\gamma_{r,Q1(2)}$ and $\gamma_{\phi,Q1(2)}$, the spin-photon couplings of both DQD's, which we assume to be identical, and an amplitude and offset of the readout signal. All other inputs to the model are experimentally measured. Explicitly, these are the charge dispersive shifts $\chi_{i}$, the resonator frequency and linewidth $\omega_r$ and $\kappa_r$, and the qubit frequencies $\omega_{Qi}$. \Cref{fig:fullmodel_fits} shows the fits of the full model to the exchange oscillations reported in the main text. The described measured respective input parameters for each panel can be seen in \cref{tab:parameter overview}.\par 

Note that here the $g_s$ values fitted from \Cref{fig:fullmodel_fits}.a-b and e-f are very close to those estimated for \Cref{fig:3} and \Cref{fig:4} using the input-output theory whereas the $g_s$ values fitted from \Cref{fig:fullmodel_fits}.c-d are slightly different. We can think of several possible contributions to such deviations. First, the magnetic field gradient is needed for the estimation based on input-output theory, but it can only be simulated and thus the number might be inaccurate. Second, in the fitted model, we have assumed that the qubits are perfectly at the same frequency. However, due to the slow electric drift in the device and consequentially the limited accuracy in the calibration, there could be a small difference in their frequencies which contributes to slightly different oscillation frequencies. Third, the slow drift can also impact the tunnel couplings and thus directly affect the $g_s$ values. Finally, when the external magnetic field is changed, the magnetic field gradient from the micromagnet can also be somewhat different, which directly affects the spin-charge hybridization and thus the $g_s$. 

\subsection*{Fidelity estimation}
We estimate the fidelity of the two-qubit gate from the fit to the data based on the full model simulations. Explicitly, we compute the average gate fidelity of the two-qubit gate with respect to the iSWAP and $\sqrt{\text{iSWAP}}$ gate ignoring single-qubit phases
\begin{align}
    1-F = \underset{\theta_{Q1},\theta_{Q2}}{\text{min}}\frac{n^2-\text{tr}\left(\chi_\text{sim} \left[U(\theta_{Q1},\theta_{Q2})\otimes U(\theta_{Q1},\theta_{Q2})^\dagger\right]\right)^\dagger}{n(n+1)}.
\end{align}
Here, $U(\theta_{Q1},\theta_{Q2}) = e^{-i \theta_{Q1} \sigma_{z,1}}e^{-i \theta_{Q2} \sigma_{z,2}} U_\text{iSWAP($\sqrt{\text{iSWAP}}$)} $ is the ideal target gate up to single-qubit phases with
\begin{align}
    U_\text{iSWAP} &=
    \left(\begin{array}{cccc}
        1 & 0 & 0 & 0 \\
        0 & 0 & i & 0 \\
        0 & i & 0 & 0 \\
        0 & 0 & 0 & 1
    \end{array}\right)\\
    U_{\sqrt{\text{iSWAP}}} &= 
    \left(\begin{array}{cccc}
        1 & 0 & 0 & 0 \\
        0 & 1/\sqrt{2} & i/\sqrt{2} & 0 \\
        0 & i/\sqrt{2} & 1/\sqrt{2} & 0 \\
        0 & 0 & 0 & 1
    \end{array}\right).
\end{align} 
Note that since the simulation is performed in the laboratory frame, the single-qubit phases simply account for the Larmor precession of each qubit.

The process matrix $\chi_\text{sim}$ can be extracted from the simulations using standard quantum process tomography techniques. Explicitly, we solve
\begin{align}
    \ket{\rho_{\text{red},ij}(t_g)} = \chi_\text{sim}(t) \ket{\rho_{\text{red},ij}(0)},
\end{align}
where $\rho_{\text{red},ij}(t)=\text{tr}_{ph}(\rho_{ji}(t))$ is the $ij$-th vectorized reduced density matrix computed from the master equation by tracing out the photonic degree of freedom, for all input states $\rho_{ij} = \ket{\psi_i}\bra{\psi_j}$ with $\ket{\psi_i}$ being the standard product basis $\lbrace\ket{00},\ket{01}\ket{10},\ket{11}\rbrace$.

\clearpage

\end{document}